\documentclass[prx,showpacs,twocolumn,amsmath,amssymb,floatfix]{revtex4-1}

\usepackage{graphicx}
\usepackage{bm}
\usepackage{color}
\usepackage{hyperref}
\usepackage[applemac]{inputenc}

\begin{document} 
   
\title{Zero-energy pinning from interactions in Majorana nanowires}

\author{Fernando Domínguez$^1$, Jorge Cayao$^3$, Pablo San-Jose$^3$, Ramón Aguado$^3$, Alfredo Levy Yeyati$^1$, Elsa Prada$^2$}
\affiliation{$^1$Departamento de Física Teórica de la Materia Condensada and $^2$Departamento de Física de la Materia Condensada, Condensed Matter Physics Center (IFIMAC) and Instituto Nicolás Cabrera, Universidad Autónoma de Madrid, E-28049 Madrid, Spain\\
$^3$Instituto de Ciencia de Materiales de Madrid, Consejo Superior de Investigaciones Científicas (ICMM-CSIC), Sor Juana Inés de la Cruz 3, 28049 Madrid, Spain}
\date{\today} 

\begin{abstract}
Majorana zero modes at the boundaries of topological superconductors are charge-neutral, an equal superposition of electrons and holes. This ideal situation is, however, hard to achieve in physical implementations, such as proximitised semiconducting nanowires of realistic length. In such systems Majorana overlaps are unavoidable and lead to their hybridisation into \emph{charged} Bogoliubov quasiparticles of finite energy which, unlike true zero modes, are affected by electronic interactions. We here demonstrate that these interactions, particularly with bound charges in the dielectric surroundings, drastically change the non-interacting paradigm. Remarkably, interactions may completely suppress Majorana hybridisation around parity crossings, where the total charge in the nanowire changes.  This effect, dubbed zero-energy pinning, stabilises Majoranas back to zero energy and charge, and leads to electronically incompressible parameter regions wherein Majoranas remain insensitive to local perturbations, despite their overlap.   
\end{abstract}

\maketitle
\section{Introduction}
Since the early experimental efforts towards the generation and characterisation of Majorana zero modes (MZMs) in nanowires \cite{Mourik:S12,Deng:NL12,Das:NP12,Churchill:PRB13,Lee:NN14}, remarkable progress has been accomplished \cite{Chang:NN15,Albrecht:N16,Zhang:16}.
Cleaner devices, with longer mean free paths and much more robust induced superconductivity, are now available. Samples of this quality are expected to develop
an unambiguous, topologically non-trivial superconducting phase hosting MZMs \cite{Kitaev:PU01}. 
Owing to their non-Abelian statistics, braiding operations of MZMs are topologically protected and hold promise as the basis of fault-tolerant quantum computers \cite{Nayak:RMP08,Sarma:NQI15}.

\begin{figure}
   \centering 
   \includegraphics[width=\columnwidth]{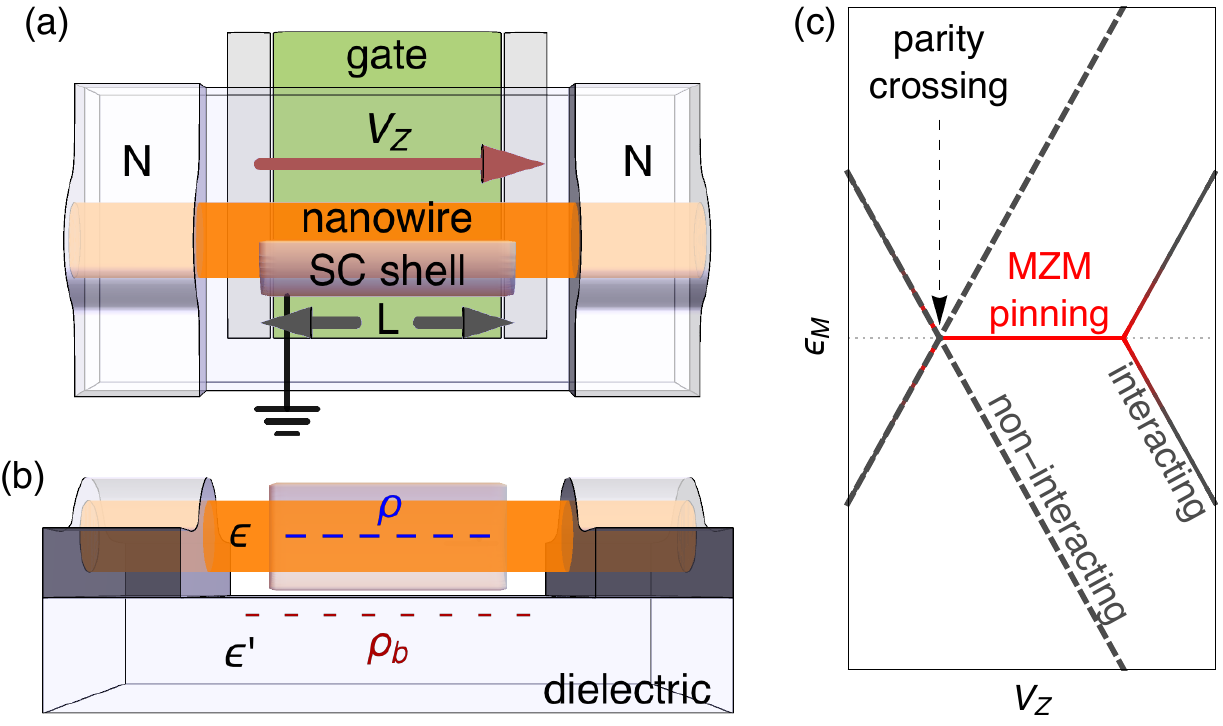}
   \caption{(a) Proximitised Rashba nanowire of length $L$ with gate-tunable Fermi energy $\mu$ and under a parallel Zeeman field $V_Z$. (b) Bound charges $\rho_b$ arise in the dielectric surroundings, which interact with free charges $\rho$ in the nanowire. (c) Sketch of the interaction-induced pinning, discussed in this work, of Majoranas at zero energy around parity crossings. This pinning provides a natural explanation to hitherto unresolved experimental features in Majorana wires (see e. g. Fig. 3c in Ref. \citenum{Albrecht:N16}).}
   \label{fig:sketch}
\end{figure}

Majoranas are topologically protected against local fluctuations in the environment inasmuch as they  {do not overlap spatially}. However, deviations from this stringent condition always occur in realistic samples of finite length $L$, see Fig. \ref{fig:sketch}a. In this case, the two MZMs at opposite ends of the wire overlap and become a charged Bogoliubov quasiparticle of finite energy $\epsilon_M$ and charge $Q_M<e$ {(assuming a macroscopic and/or grounded parent superconductor) \footnote{The case of a floating superconductor with charging energy has been studied in e.g. Refs. \citenum{Fu:PRL10,Hutzen:PRL12}.}}. 
While these deviations are expected to be exponentially small, their importance of course depends on the spatial extension of the Majoranas $\xi_M$ as compared to $L$ (or the wire's mean free path, whichever is smaller), since both  $\epsilon_M$ and $Q_M$ decrease as $\sim e^{-L/\xi_M}$.

Even in the longer, $L\sim 1\mu$m nanowires experimentally studied so far, deviations from true MZM behaviour are not negligible, {since typically $\xi_M$ is in the hundreds of nanometers}  \footnote{In the weak SO regime, which corresponds to typical SO lengths for InSb nanowires of $L_\mathrm{SO}\approx 250$nm (SO coupling $\alpha\sim 0.2$ eV\AA), the Majorana localization length is $\xi_M\sim L_\mathrm{SO}$, and furthermore increases with Zeeman field \cite{Cayao:PRB15}. The condition $L\gg\xi_M$ cannot be fulfilled as the Zeeman field exceeds the critical value above the topological transition. If, on the other hand, we assume stronger SO couplings, $\xi_M$ saturates to $\xi_M\sim\alpha/\Delta$ \cite{Klinovaja:PRB12,Mishmash:PRB16}, which, for typical values of $\Delta$, is still much larger than the SO length. For example, assuming a SO coupling ten times larger than before, $\alpha\sim 2$ eV\AA, and a proximity gap $\Delta=0.5$meV (of the order of the experimental one in \cite{Zhang:16}), the Majorana localisation length is $\xi_M\sim 400$nm.  The $L\gg\xi_M$  limit would therefore need much longer wires than the ones studied so far, which are in the range $L\lesssim 1\mu$m}. This theoretical expectation is in stark contrast with many experiments reporting surprisingly robust zero-bias anomalies (see e.g. Ref. \citenum{Zhang:16}). Within a non-interacting picture \cite{Lutchyn:PRL10,Oreg:PRL10}, the only solution to this conflict is to assume values of the spin-orbit coupling much larger than the ones estimated for InAs or InSb NWs \cite{Mishmash:PRB16}. Apart from the exponential suppression, $\epsilon_M$ is expected to cross zero energy (parity crossings where $Q_M$ changes) in an oscillatory fashion \cite{Prada:PRB12,Das-Sarma:PRB12,Rainis:PRB13} as a function of magnetic field, chemical potential or length.  Both the exponential suppression and the oscillatory behaviour of $\epsilon_M$ have recently been demonstrated experimentally \cite{Albrecht:N16}. However, these experiments report on unexplained features in rather short wires with $L\sim\xi_M$ in the form of parity crossings that extend across a finite range of magnetic field (instead of point-like zero energy crossings).

In this work, we present an alternative to the non-interacting view that provides an explanation to the above unresolved issues. By considering electronic interactions with the electrostatic environment \cite{Vuik:NJP16}, we demonstrate that zero-energy crossings originating from the oscillatory splitting of overlapping MZMs are spontaneously stabilised into extended regions in parameter space wherein Majoranas become pinned to zero energy (Fig. \ref{fig:sketch}c).  {Our results show that the absence of Majorana splittings, which is commonly identified with non-overlapping Majoranas with topological protection, can occur despite the Majorana overlap in nanowires of finite length}.

The interactions involved in Majorana zero-energy pinning are not intrinsic to the wire \cite{Stoudenmire:PRB11,Lutchyn:PRB11,Sela:PRB11,Gangadharaiah:PRL11,Fidkowski:PRB12,Das-Sarma:PRB12,Manolescu:JPCM14, Ghazaryan:PRB15,Xu:PLA16}, but rather extrinsic, between electrons that enter the wire and bound charges in the dielectric environment that arise in response. Our results suggest that such electronic interactions provide a powerful mechanism to stabilise {Majorana-based} qubits in realistic nanowires, and may account for the hitherto unexplained experimental features \cite{Zhang:16, Albrecht:N16}.

\section{Results}

The central idea behind the zero-energy pinning phenomenon is the emergence of repulsive self-interactions through image charges in the dielectric medium. Before presenting full microscopic calculations, we first illustrate the mechanism with a toy example. Assume a quantum system with a single-particle state $\psi_M$ that carries an electric charge $Q_M$. In the absence of interactions its energy is $\epsilon_M$. The Hamiltonian, including a `self-interaction' term, takes the form $H=\epsilon_M \psi_M^\dagger\psi_M+V_bQ_M\langle\psi_M^\dagger\psi_M\rangle\psi_M^\dagger\psi_M$, with $eV_b$ an energy scale. Thus, in the presence of interactions, the effective level $\tilde\epsilon_M$ is the self-consistent solution to $\tilde \epsilon_M=\epsilon_M+V_b Q_Mf(\tilde\epsilon_M)$, where $f(\epsilon)=1/(e^{\epsilon/k_BT}+1)$ is the Fermi function. As $\epsilon_M$ is externally tuned by a parameter $V_Z$ to cross the Fermi energy (defined as zero), the resulting electron and hole energies $\pm\tilde\epsilon_M$ with interactions (solid curves, $V_b>0$ [repulsive]) and without (dashed, $V_b=0$) are of the form shown in Fig. \ref{fig:sketch}c for $T\to 0$. We see that the $V_b=0$ parity crossing at $\epsilon_M(V_Z)=0$ transforms, for repulsive $V_b>0$, into a finite plateau (in red), wherein $\tilde\epsilon_M$ becomes pinned to zero within a finite range of $V_Z$, $0<\epsilon_M(V_Z)<V_bQ_M$. The pinning plateau is electronically incompressible, {and is the result of thermal equilibrium combined with the self-interaction energy cost of occupying state $\psi_M$}.

The above toy model for pinning presents a fundamental question when considering Majoranas in the role of the $\psi_M$ state: how do physical self-interactions arise?
It is clear that self-interactions of the form $\langle\psi_M^\dagger\psi_M\rangle\psi_M^\dagger\psi_M$ are unphysical in an isolated quantum system.
In a generic basis, a direct (intrinsic) charge-charge interaction of the form $H_C=\sum_{ij}\psi_i^\dagger\psi_i V_{ij}\psi^\dagger_j\psi_j$ produces self-interaction of eigenstates when treated at the Hartree-level, but this is cancelled by the Fock correction \cite{Cowan:PR67,Perdew:PRB81,Anisimov:00}.

However, if the electronic system is immersed in a dielectric medium, a bound charge density $\rho_b(\bm r)$ may appear, in response to electric charges $\hat \rho(\bm r)$ in the system, at interfaces where the dielectric constant $\epsilon(\bm r)$ changes, see Fig. \ref{fig:sketch}b. These $\rho_b(\bm r)$ generate an electrostatic potential $\phi(\bm r)$ that acts back onto $\hat \rho(\bm r)$, such that the total potential $\phi_\mathrm{tot}(\bm r)=\phi_\mathrm{sys}(\bm r)+\phi(\bm r)$ satisfies the inhomogeneous Poisson equation $\bm\nabla\cdot\left[\epsilon(\bm r)\bm\nabla\phi_\mathrm{tot}(\bm r)\right]=-4\pi\langle\hat\rho(\bm r)\rangle$, and $\phi_\mathrm{sys}(\bm r)$ is the potential for an infinite system with a uniform $\epsilon$ (without bound charges). The resulting (extrinsic) interaction between the system's $\hat\rho(\bm r)$ and bound charges $\rho_b(\bm r')$ then takes the form of a Hartee-like physical self-interaction, $\int \phi(\bm r)\hat\rho(\bm r)=\int V_b(\bm r',\bm r)\langle\hat\rho(\bm r')\rangle\hat\rho(\bm r)$, where $V_b$ depends on the actual device geometry. Note that no Fock-like correction should be included here, since this is a purely classical effect: bound charges are located outside the nanowire and
can be distinguished from the free charges. Thus, an effective Hamiltonian similar to the toy model above becomes relevant. Interaction with bound charges should be expected to produce zero-energy pinning of a quantum state as long as they are repulsive (i.e. if $\rho_b$ and $\rho$ have the same sign). The latter condition is satisfied if the dielectric environment has a smaller dielectric constant than the nanowire, the typical situation in most experiments (e.g. InSb or InAs nanowires on a SiO${}_2$ substrate, {see the Discussion section for further details}). Such self-interactions are well known in the context of molecular junctions \cite{Kaasbjerg:NL08,Kaasbjerg:PRB11} but, to our knowledge, their implications have not been explored in the context of Majorana wires.

\begin{figure}
   \centering 
   \includegraphics[width=\columnwidth]{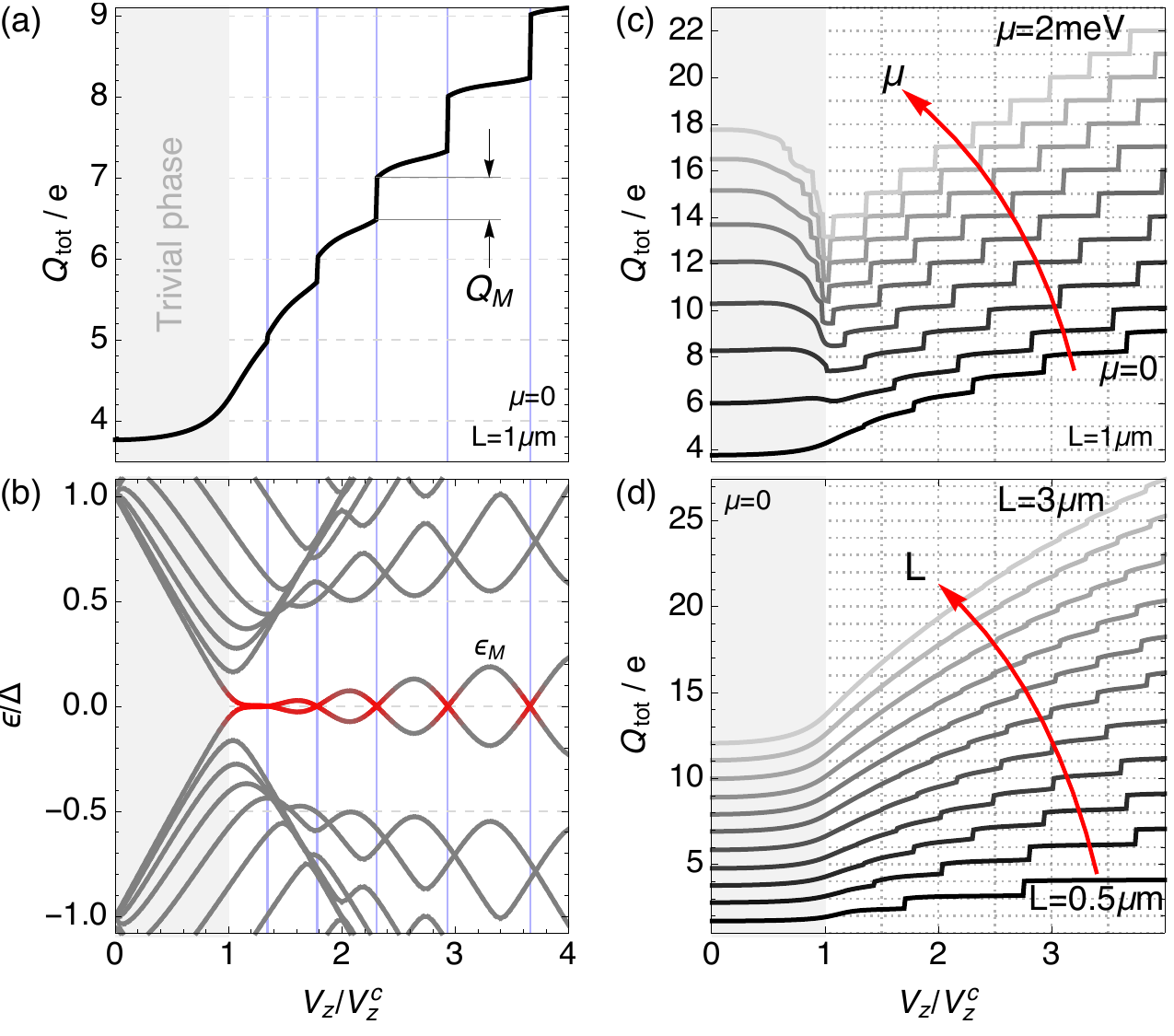}
   \caption{(a) Total charge $Q_\mathrm{tot}$ in a non-interacting topological nanowire versus $V_Z$. For $V_Z>V_Z^c$, $Q_\mathrm{tot}$ increases in jumps of magnitude $Q_M<e$ at parity crossings, shown in red in (b). (c,d) Charge jumps become more (less) quantised as $\mu$ ($L$) increases.}
   \label{fig:charging}
\end{figure}

A second relevant question concerns the charge $Q_M$ associated to Majoranas. MZMs in infinite superconducting systems are charge-neutral, an equal superposition of electrons and holes. As such, it might come as a surprise that electronic interactions, which are sensitive to electronic charge, would have an effect on Majoranas. It was demonstrated \cite{Ben-Shach:PRB15,Lin:PRB12}, however, that the state resulting from the hybridisation of two overlapping MZMs, each located at an end of a nanowire of length $L$, is not charge-neutral, but rather defines a charged Bogoliubov quasiparticle of finite energy $\epsilon_M$ and charge $Q_M<e$, with typical $\epsilon_M$ and $Q_M$ decreasing exponentially as $\sim e^{-L/\xi_M}$. We now analyse this phenomenology in detail within a non-interacting model \cite{Oreg:PRL10,Lutchyn:PRL10} of a proximitised InSb semiconducting nanowire, oriented along the $x$-direction, of length $L=1\mu$m and subjected to a parallel Zeeman field $V_Z=\frac{1}{2}g\mu_B B$ ($g$ is the g-factor, $\mu_B$ is the Bohr magneton and $B$ is the magnetic field). The continuum Hamiltonian without superconductivity reads $H_0=\frac{\hbar^2k^2}{2m}+\alpha \bm{\sigma}_y k + V_Z \bm{\sigma}_x$, where $\hbar k$ is the momentum along the wire, $\bm \sigma$ is the spin, the InSb spin-orbit coupling is $\alpha=0.2$ eV \AA, and the effective mass is $m=0.015 m_e$. We include the induced superconducting pairing $\Delta=0.5$ meV in the second-quantised Nambu representation,
\begin{eqnarray*}\label{H0}
H=\frac{1}{2}\int dx\,\left(\Psi^\dagger(x),\Psi(x)\right)\!
\left(\begin{array}{cc}
H_0-\mu & -i\bm{\sigma}_y\Delta^* \\
i\bm{\sigma}_y\Delta & \mu-H_0^*
\end{array}\right)\!
\left(\!\begin{array}{c}
\Psi(x)\\ \Psi^\dagger(x)
\end{array}\!\right),
\end{eqnarray*}
{where $\Psi(x)=(\Psi_\uparrow(x),\Psi_\downarrow(x))$ is the electron field, so that the charge density reads $\hat{\rho}(x)=e\sum_\sigma\Psi^\dagger_\sigma(x)\Psi_\sigma(x)$.}

{Figures \ref{fig:charging}(a,b) show the total charge in the nanowire  $Q_\mathrm{tot}=\int_0^L dx \langle\hat\rho(x)\rangle$ at zero temperature and the low-energy spectrum as a function of $V_Z$, respectively}. Any given Bogoliubov quasiparticle eigenstate
\begin{equation}
\psi_n=\int_0^L dx\left[ u_n(x)\Psi(x)+v_n(x)\Psi^\dagger(x)\right]
\end{equation}
{in the Nambu spectrum with energy $\epsilon_n>0$ contributes to $Q_\mathrm{tot}=\sum_n Q_n$} with a charge 
\begin{equation}
Q_n=e\int_0^L dx\left\{f(\epsilon_n)|u_n(x)|^2+[1-f(\epsilon_n)]|v_n(x)|^2\right\},
\end{equation}
where $f(\epsilon_n)$ is its occupation probability. For $V_Z>V_Z^{c}=\sqrt{\Delta^2+\mu^2}$ the nanowire becomes topologically non-trivial and, for $L\to\infty$, develops MZMs at each end $\gamma_{L,R}=\gamma_{L,R}^\dagger$ {with $u_{L,R}(x)=v^*_{L,R}(x)$. These states have zero charge, since changing their occupation does not change $Q_\mathrm{tot}$. For finite $L$, the MZMs overlap and are no longer eigenstates,} but hybridise into two {special $n=\pm 1$} {Nambu eigenstates $\psi_1=\psi_M=(\gamma_L+ i\gamma_R)/2$ and $\psi_{-1}=\psi^\dagger_M=(\gamma_L-i\gamma_R)/2$}, whose energies $\pm\epsilon_M$ oscillate around zero as a function of $V_Z$, $\mu$ or $L$. Each time $\epsilon_M$ crosses zero in Fig. \ref{fig:charging}b (parity crossing), the {zero-temperature occupation of the $\psi_M$ quasiparticle at equilibrium changes abruptly, and the total charge of the wire experiences a non-quantised jump $Q_M=|Q_1-Q_{-1}|$, as shown in Fig. \ref{fig:charging}a. Thus, the charge $Q_M$ of the non-local Majorana fermion $\psi_M$} is non-zero despite it being a superposition of the two neutral MZMs, and is distributed almost uniformly along the wire. It reads 
\begin{eqnarray}
Q_M&=&\frac{e}{4}\int_0^L dx \left[\left|u_L^*(x)+iu_R^*(x)\right|^2-\left|u_L^*(x)-iu_R^*(x)\right|^2\right]\nonumber\\
&=&e\,\int_0^L dx |u_L(x)u_R(x)|,
\end{eqnarray}
i.e. $Q_M$ is the spatial overlap of the two Majoranas {(here we have chosen $u_{L}$ real and $u_{R}$ imaginary without loss of generality)}. As we discussed in the introduction, the spatial extension of MZMs $\xi_M$ is typically larger than the spin-orbit length $L_\mathrm{SO}=\hbar^2/\alpha m$ \cite{Klinovaja:PRB12}, which is seldom smaller than a few hundred nanometers ($L_\mathrm{SO}=254$ nm here). Thus, even in rather long $L\sim 1\mu$m non-interacting topological nanowires, the overlap of the Majoranas is sizeable, and results in large $Q_M\sim e$ charge jumps, see Fig. \ref{fig:charging}a.

\begin{figure}
   \centering 
   \includegraphics[width=\columnwidth]{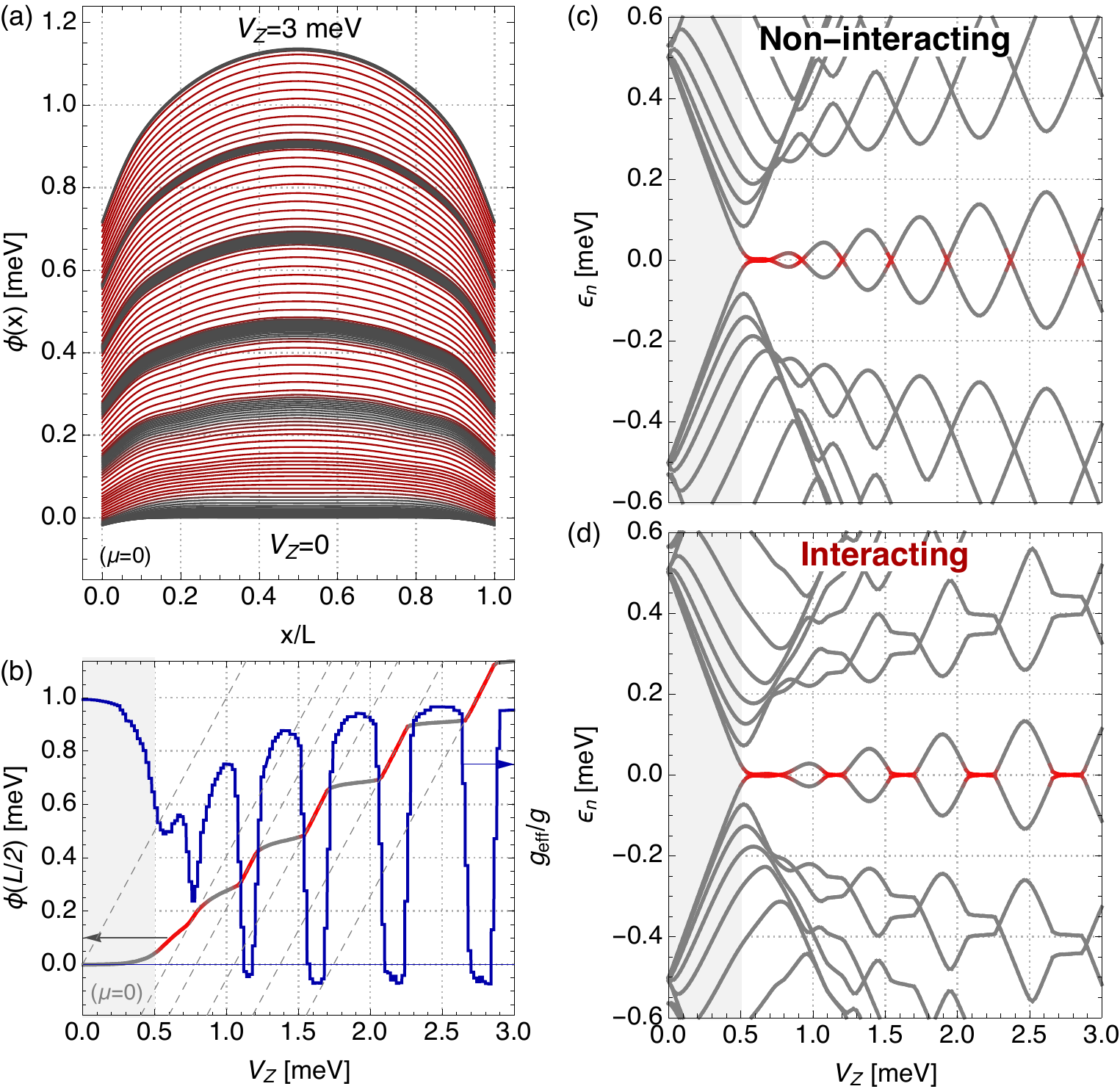}
   \caption{(a) Screening potential $\phi(x)$ along a proximitised InSb nanowire of radius $R=50$ nm and length $L=1\mu$m for increasing Zeeman field $V_Z$. We take $\Delta=0.5$ meV and $\mu=0$. Red (gray) curves correspond to an incompresible (compressible) regime. (b) Increase of the potential at the center of the wire $\phi(L/2)$ with $V_Z$, and the corresponding effective g-factor $g_\mathrm{eff}=g\left[1-\partial\phi(L/2)/\partial V_Z\right]$ (blue line). Intervals in red correspond to almost perfect screening of $V_Z$ ($g_\mathrm{eff}\approx 0$). (c,d) Non-interacting and interacting spectra of the nanowire. Note that perfect Zeeman screening by interactions is correlated with zero-energy pinnings of Majorana bound states (extended zero modes in red). An equivalent simulation to (d) including also intrinsic interactions in the nanowire is shown in Fig. 1 of the Appendix.}
   \label{fig:pinning}
\end{figure}

\begin{figure}
   \centering 
   \includegraphics[width=\columnwidth]{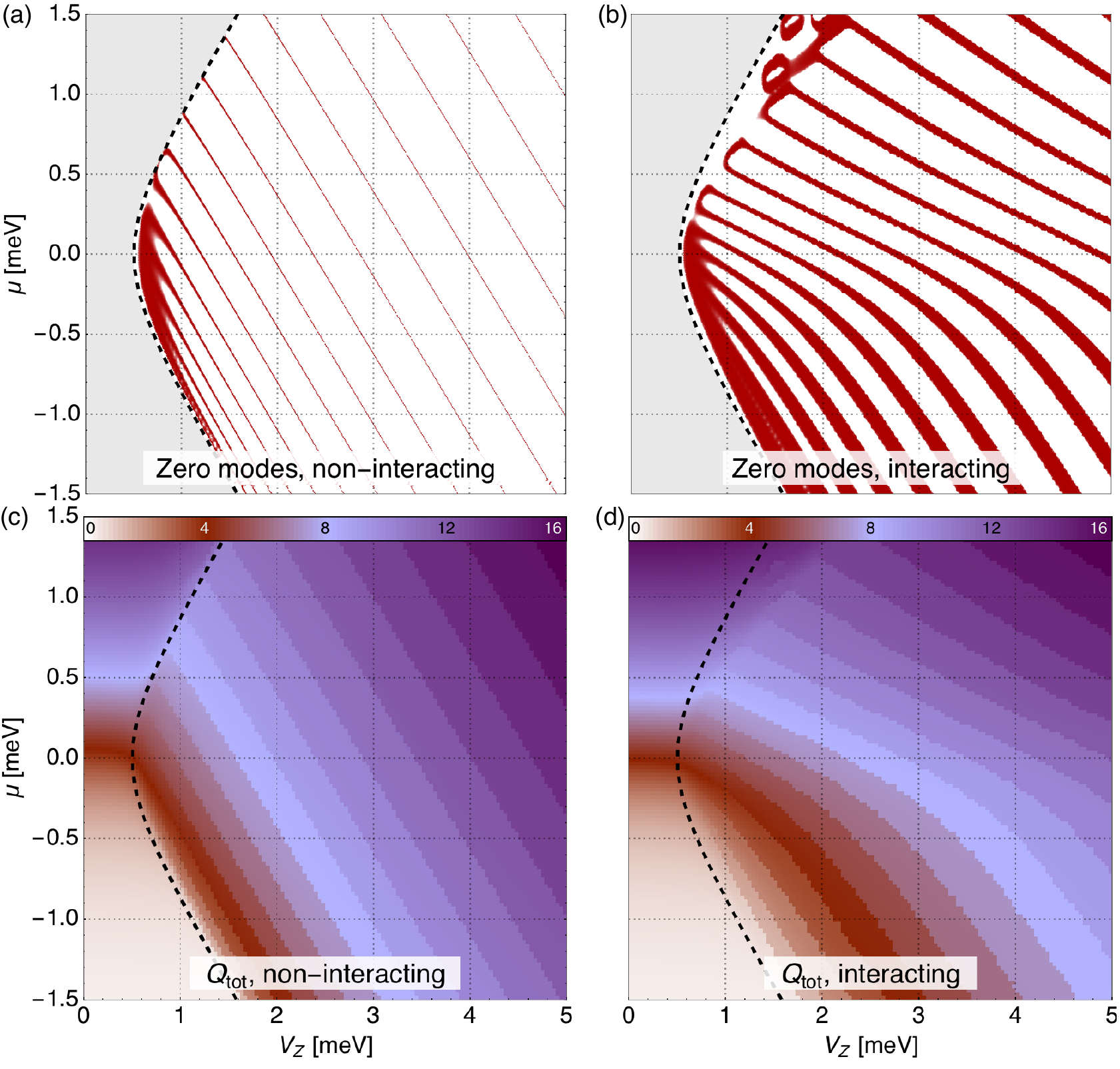}
   \caption{Topological phase diagram of a non-interacting (a,c) and interacting (b,d), finite-length nanowire, see Fig. \ref{fig:pinning} for parameters. The dashed line corresponds to the non-interacting topological transition $V_Z^c=\sqrt{\Delta^2+\mu^2}$. Red areas in (a,b) represent zero-energy modes (with energy below 10 $\mu$eV at $T=10$mK), which correspond to narrow parity crossings from hybridized MZMs in (a), and extended incompressible regions of MZMs pinned at zero energy in (b). The total charge in the nanowire, panels (c,d), increases by $Q_M$ jumps at each zero energy crossing.}
   \label{fig:phasediagram}
\end{figure}

Since parity-crossings introduce a charge $Q_M$ into the system, they should be expected to give rise, in a repulsive dielectric environment, to bound charges of the same sign and hence to repulsive self-interactions and to zero-energy pinning of the corresponding MZMs. To demonstrate this effect we now add interactions to the above InSb Rashba nanowire model. Following the preceding discussion, we include only interactions with bound charges in the dielectric medium, described at a self-consistent Poisson mean field level. (The addition of intrinsic interactions do not change the essential results, and are analysed in the Appendix.)
We replace $H_0$ in the non-interacting nanowire Hamiltonian $H$ above with $H_0 + e\phi(x)$, where the self-consistent potential created by bound charges is written as
\begin{equation}\label{phi}
\phi(x)=\int dx'\,V_b(x,x')\langle \hat\rho(x')\rangle.
\end{equation}
We assume a simple geometry and a Coulomb interaction with image charges of the form $V_b(x,x')\approx \frac{\epsilon-\epsilon'}{\epsilon+\epsilon'}/(4\pi\epsilon_0\epsilon\sqrt{(2R)^2+|x-x'|^2})$, where $\epsilon=17.7$ is the InSb dielectric constant, $\epsilon'=3.9$ is that of a SiO$_2$ substrate, and $R=50$ nm is the nanowire radius (see the Appendix for a derivation). The results are qualitatively independent of $V_b$, as long as it is repulsive ($\epsilon>\epsilon'$). We reabsorb the $V_Z=0$ value of the potential $\phi(L/2)$ at the center of the nanowire into $\mu$, so that the actual Fermi energy at zero magnetic field is equal to $\mu$ with and without interactions.

The potential $\phi(x)$ is solved numerically by self-consistent iteration (see the Appendix for analytical results). The solution for $\mu=0$ is shown in Fig. \ref{fig:pinning}a as a function of $V_Z$, while $\phi(L/2)$ evaluated at the center of the nanowire is shown in Fig. \ref{fig:pinning}b.  The $V_Z$-dependent $\phi(L/2)$ transforms the simple non-interacting Fermi energy $\varepsilon_F=\mu+V_Z$ into a non-trivial $\varepsilon_F^\mathrm{eff}=\mu+V_Z-\phi(L/2)$ at the center of the nanowire, changing also the $V_Z$-dependence of the spectrum as a result.
Panels (c) and (d) show the corresponding low energy spectrum without and with interactions, respectively. For $V_Z>V_Z^c=0.5$ meV, parity crossings emerge in the non-interacting spectrum that inject a finite charge $Q_M$ into the system (Fig. \ref{fig:charging}a). At each of these points, the interactions with bound charges conspire to suppress the charging, giving rise to finite intervals of $V_Z$ with pinned zero-energy modes in place of parity crossings (panel [d]). These are the result of a rapid increase in the overall self-consistent $\phi(x)$ within each interval (red curves in panels [a,b]), which screens the Zeeman field in $\varepsilon_F^\mathrm{eff}$ almost completely, making the system incompressible $\partial\varepsilon_F^\mathrm{eff}/\partial V_Z\approx 0$. This intermittent incompressibility is visible in Fig. \ref{fig:pinning}b as a slope $\partial\phi(L/2)/\partial{V_Z}\approx 1$. The electronic compressibility against variations of $V_Z$ may be quantified by the effective g-factor $g_\mathrm{eff}=g\,\partial\varepsilon_F^\mathrm{eff}/\partial V_Z=g\left[1-\partial\phi(L/2)/\partial{V_Z}\right]$, in blue.  Deviations from exact zero-energy pinning and perfect incompressibility $g_\mathrm{eff}=0$ may arise from finite temperature, $T=10$mK in these simulations, or finite decay rate into the reservoirs, neglected here. 

In the above simulation the chemical potential was taken as $\mu=0$. A similar phenomenology persist also at different electronic densities. Figure \ref{fig:phasediagram} shows the incompressible pinned regions (panels [a,b], in red) and total charge $Q_\mathrm{tot}$ in the nanowire (panels [c,d]) across the full $\mu-V_Z$ parameter space, with and without interactions. When interactions are switched on, zero-measure parity crossings grow into extended areas of pinned MZMs, always separated by areas with finite MZM hybridisation $|\tilde\epsilon_M|>0$ (white). Along a given line within each incompressible area, the charge in the nanowire jumps by a finite, almost constant $Q_M$.  

We note that, apart from stabilising MZMs, interactions induce a change around $\mu=0$ in the slope $s$ of incompressible regions in the $\mu-V_Z$ plane, with $s=-1$ for $\mu<0$ like in the non-interacting case, and $-1<s<0$ for $\mu>0$. The incompressible regions are contours of constant $\varepsilon_F^\mathrm{eff}$~\cite{Klinovaja:PRB12}, so that $s=
-g\kappa_\mathrm{eff}/g_\mathrm{eff}$ is a ratio between system's $\mu$-compressibility $\kappa_\mathrm{eff}=\partial\varepsilon_F^\mathrm{eff}/\partial \mu$, and the $V_Z$-compressibility $g_\mathrm{eff}/g=\partial\varepsilon_F^\mathrm{eff}/\partial V_Z$ discussed above. While both are equal for $\mu<0$, $\kappa_\mathrm{eff}$ becomes suppressed for $\mu>0$, so that $s\approx-1/(1+\sqrt{v/\mu})$ for a constant $v\propto V_b^2$ related to the interaction strength (see Supplementary Information). 

\section{Discussion}

We have discussed a generic mechanism whereby electrostatic interactions with the dielectric surroundings stabilise zero energy modes in a finite length Majorana nanowire.  This zero-energy pinning effect is the result, within a self-consistent mean field, of a repulsive self-interaction of nanowire electrons that arises when its dielectric environment has a smaller dielectric constant than the nanowire itself. {In a more general situation in which the charge screening by the normal contacts and the parent superconductor are also considered (a situation not presented here for simplicity, but with qualitatively similar results), the relevant quantity becomes the difference between the total electrostatic energy of a charge inside and outside the proximitized nanowire. This energy difference becomes positive, and hence self-interactions are repulsive, if charge screening is \emph{reduced} upon entering the proximitized region. Such is the natural situation for nanowires with partial superconducting shells but full metallic covering at the contacts. The addition of charge into the nanowire calculated self-consistently then leads to pinning}. Related dielectric-induced changes in addition energies have been discussed in molecular single electron transistors \cite{Kaasbjerg:NL08}. Zero-energy pinning cannot be captured by solving the electrostatic problem for infinite nanowires \cite{Vuik:NJP16}, since it is necessary to take into account the electrostatic energy cost of adding charge at each parity crossing. Pinning, moreover, does not require a single channel regime. It also operates for any odd number of open channels \cite{Potter:PRB11,Stanescu:PRB11,San-Jose:PRL14}, as the physics of parity crossings is similar.

Within the stabilised regions in parameter space with pinned Majoranas, the system becomes electronically incompressible \footnote{A related phenomenon of global incompressibility, albeit unconnected to zero-energy pinning and parity crossings, was discussed in Majorana nanowires in the limit of strong intrinsic interactions \cite{Das-Sarma:PRB12}.}. As a result, potential fluctuations $\delta\mu$ from the environment or fluctuations $\delta V_Z$ in the applied Zeeman field, become screened out by interactions. {This should remain true even for spatially non-uniform perturbations. As a result, the pinning effect could potentially be exploited to protect realistic Majorana-based qubits against environmental noise.} We anticipate that, by carefully engineering the dielectric surroundings of finite-length Majorana nanowires, one could exploit the electronic incompressibility of pinned regions to {replace, at least partially, the topological protection of MZMs against decoherence which is lost by their overlap}. In this respect it is important to emphasise that pinned MZMs still overlap spatially, but become locked into {degenerate parity eigenstates, regardless of any local perturbation, as long as thermal equilibrium is preserved. We should stress that, despite the common misconception in large part of the literature that identifies robust zero Majorana splitting with topological protection, it is non-locality what ultimately renders fermionic parity qubits immune to local noise. While there is a strong resemblance, the resilience to arbitrary perturbations discussed in this work is different from topological protection. The implications {of pinning} for the decoherence and relaxation times of parity-conserving Majorana qubits, {such as} e.g. the $|00\rangle$ and $|11\rangle$ even-parity states of a four-Majorana setup, are non-trivial and should be the subject of future work.}

The zero-energy pinning mechanism {described in this work} is generic, and is the result of the electrostatic energy cost from the interaction of the finite charge $Q_M$ added to the nanowire at parity crossings and image charges in the dielectric environment. Given the generality of the mechanism, we speculate that the ideas discussed here are also relevant in other contexts, including parity crossings of Shiba states in non-topological superconductors. 

\section{Methods} 

All the numerical results were computed within a self-consistent mean-field treatment of interactions in a tight-binding model for the semiconducting nanowire. We used the  MathQ framework \cite{MathQ} on a spatial discretisation of the nanowire model $H_0=\frac{\hbar^2k^2}{2m}+\alpha \bm{\sigma}_y k + V_Z \bm{\sigma}_x$. The  lattice spacing used is 10 nm. The mean field self-consistency condition, encoded in Eq. (\ref{phi}), is achieved by iteration with an adaptive update coefficient. This is required at low temperatures to achieve convergence around pinned regions.

\acknowledgements

We thank K. Flensberg for illuminating discussions. We acknowledge financial support from the Spanish Ministry of Economy and Competitiveness through the Ram\'on y Cajal program RYC-2013-14645 and RYC-2011-09345, grant Nos. FIS2012-33521, FIS2013-47328-C2-1-P, FIS2014-55486-P,  FIS2015-65706-P, FIS2015-64654-P (MINECO/FEDER), and the ``Mar\'ia de Maeztu'' Programme for Units of Excellence in R\&D (MDM-2014-0377).

\appendix

\section{Electron-electron interactions in a Majorana nanowire}

We study the electronic interactions between charges in a Majorana nanowire, like the one considered in the main text, when this nanowire is surrounded by a dielectric medium. The interactions can be divided into two contributions: the so-called \emph{intrinsic} term, that accounts for the direct interaction between electrons of density $\hat\rho$ within the wire, and the \emph{extrinsic} term, resulting from the interaction of the $\hat \rho$ charge density with bound polarization charges $\rho_b$, that emerge at dielectric boundaries as a response to $\hat \rho$ itself. We analyse them in the two following subsections.

\subsection{Intrinsic potential}

Consider the nanowire along the $\hat x$ axis as an isolated quantum system embedded in a uniform dielectric medium with constant $\epsilon$. No bound charges arise anywhere in this configuration. We thus analyse the effect of direct electronic interactions between electrons in the nanowire. We do this in two steps. We first consider Coulomb interactions screened, in the quasi-static Thomas-Fermi limit, by mobile charges along the nanowire. This corresponds to dressing the electromagnetic field propagator, which then yields a short range interaction,
\[
V_{TF}(x)=\frac{e^{-|x|/\lambda_{TF}}}{4\pi\epsilon_0\epsilon |x|}
\]
The Thomas-Fermi length $\lambda_{TF}$ is proportional to the inverse nanowire charge density and should thus be much larger than in typical metals. A more precise potential taking into account the finite radius $R$ of the charge density across the nanowire section can be derived \cite{Giuliani:05}, that effectively introduces a cutoff at distances $|x|<R$
\[
V_{TF}(x)=\frac{\sqrt{\pi}}{4\pi\epsilon\epsilon_0 R}e^{|x|^2/R^2-|x|/\lambda_{TF}}\mathrm{Erfc}(|x|/R)
\]
The Hamiltonian for quasiparticles in the nanowire becomes $H=H_0+H_{int}$, where 
\[
H_{int}=\int dx dx' \hat\rho(x)V_{TF}(x-x')\hat \rho(x'),
\]
and $H_0$ for the non-interacting nanowire is given in the main text. 

To treat these interactions, we next take a mean field approach, which is equivalent to dressing the electronic propagator. One must take care to include both Hartree and Fock terms in the mean field decoupling,
\begin{eqnarray*}
H_{int}^{MF}&=&\int dx dx' \, V_{TF}(x-x')\\
&&\times\left[\hat\rho(x)\langle\hat \rho(x')\rangle-\psi^\dagger(x)\psi(x')\langle\psi^\dagger(x')\psi(x)\rangle\right]
\end{eqnarray*}

The second term (Fock) ensures that non-physical self-interactions introduced by the first term (Hartree) are cancelled (approximately). This approach to interactions is only meaningful in one-dimensional metals for short enough nanowires, so that the level spacing exceeds the typical interaction strength. Otherwise, in the long-nanowire limit, the Fock term leads to the opening of an unphysical mean-field gap, which signals the breakdown of the Fermi-liquid approach. The universality class in that limit becomes that of the Luttinger liquid. For the above screened interactions, and micron-long InSb nanowires, the Hartee-Fock approach to intrinsic interactions remains reasonable.

Another subtlety to be considered is that, in the proximity of a superconductor, the mean field decoupling should also include self-consistent anomalous pairing terms $\sim-\psi^\dagger(\bm r)\psi^\dagger(\bm r')\langle\psi(\bm r')\psi(\bm r)\rangle$, which we didn't include in the equation above. While their effect is negligible in relation with parity crossings, they produce a substantial suppression of the superconducting gap in the non-topological phase \cite{Gangadharaiah:PRL11,Sela:PRB11,Stoudenmire:PRB11,Ghazaryan:PRB15}.

\begin{figure}[t]
   \includegraphics[width=\columnwidth]{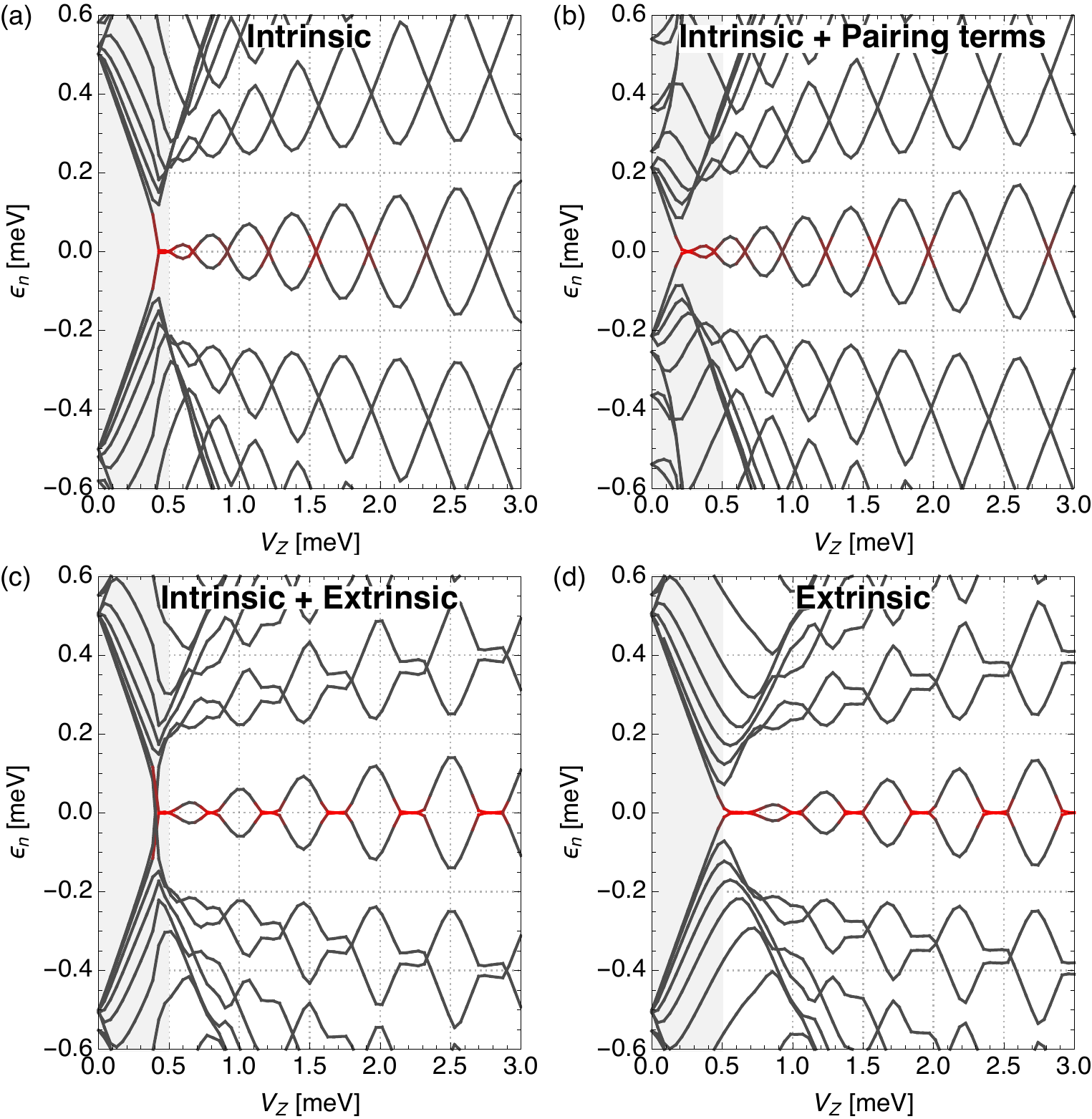} 
   \caption{Low energy spectrum of the $L=1\mu$m InSb nanowire of the main text, including interactions within the nanowire at different levels of approximation. Including only intrinsic interactions at the Hartree-Fock (a) and Hartree-Fock-Pairing level (b) does not lead to zero-energy pinning. Extrinsic interactions, either with (c) and without (d) the intrinsic Hartree-Fock contribution, leads to very similar zero-energy pinning in the topological phase (red plateaux).}
   \label{fig:extint}
\end{figure}

We now compute the low-energy spectrum as the system is driven by $V_Z$ into the topological regime, including the self-consistent $H_{int}^{MF}$. We use parameters as in the main text, and a Thomas-Fermi length $\lambda_{TF}=10$nm. The result is shown in Fig. \ref{fig:extint}a for $\mu=0$. We note that no zero-energy pinning of Majorana zero modes arises at parity crossings, as the Fock term completely cancels intrinsic self-interactions. The addition of pairing terms in the mean field decoupling does not change this fact, and merely suppresses the $V_Z=0$ gap, as mentioned above, see Fig. \ref{fig:extint}b. Naturally, it is this renormalized gap that should be fitted to the experimental observation.

We therefore see that, regarding the zero-energy pinning mechanism, there is no qualitative difference between considering only extrinsic interactions with bound charges, as done in the main text (Fig. \ref{fig:extint}d) and adding also the Hartree-Fock intrinsic interactions (Fig. \ref{fig:extint}c). The extension and position of the pinned, incompressible regions are almost unchanged. This justifies the simplified approach taken in the main text.

\subsection{Extrinsic potential}

In the preceding section we considered a nanowire embedded in an infinite, homogeneous dielectric medium, with the same dielectric constant $\epsilon$ as the nanowire itself. If the nanowire is immersed in a dielectric medium with a different constant $\epsilon'\neq \epsilon$, one must include a new source of electronic interactions. These are extrinsic to the wire, and are due to a bound charge density $\rho_b(\bm r)$  that will appear, in response to electric charges $\hat \rho(\bm r)$ in the nanowire, at interfaces where the dielectric constant $\epsilon(\bm r)$ changes. These $\rho_b(\bm r)$ generate an electrostatic potential $\phi_b(\vec r)$ that acts back onto $\hat \rho(\bm r)$, such that the total potential $\phi_\mathrm{tot}(\bm r)=\phi_\mathrm{sys}(\bm r)+\phi_b(\bm r)$ satisfies the inhomogeneous Poisson equation $-\bm\nabla\cdot\left[\epsilon(\bm r)\bm\nabla\phi_\mathrm{tot}(\bm r)\right]=4\pi\langle\hat\rho(\bm r)\rangle$, and $\phi_\mathrm{sys}(\bm r)$ is the potential for an infinite system with a uniform $\epsilon$ (without bound charges) discussed in the previous section,
\[
\phi_\mathrm{sys}(x)=\int dx V_{TF}(x-x')\langle \rho(x')\rangle.
\]
The intrinsic interaction at a Hartree level can thus be written as $\int \phi_\mathrm{sys}(x)\hat\rho(x)$, with the resulting unphysical self-interactions cancelled by the Fock term. On the other hand, the extrinsic interaction between the system's $\hat\rho(\bm r)$ and bound charges $\rho_b(\bm r')$ then takes the form of a Hartee-like physical self-interaction, $\int \phi_b(\bm r)\hat\rho(\bm r)=\int V_b(\bm r',\bm r)\langle\hat\rho(\bm r')\rangle\hat\rho(\bm r)$ for a certain effective interaction $V_b$. No Fock term should be included here, as $\rho_b$ are bound charges distinct from the free charges $\rho$, and are located outside the wire.

We now derive an approximate expression for the effective interaction $V_b$. An analytical expression for $V_b$ in a finite-length cylindrical nanowire requires solving the density $\rho_b(\bm r)$ in closed form. This problem has no known solution. Even for an infinite cylinder, the known expressions are rather unwieldy. A precise computation including all the elements present in the electromagnetic environment of experimental nanowires requires a fully three-dimensional numerical simulation. The most important aspect of the solution, however, is the short- and long-distance behaviours of $V_b(x)$, which do not crucially depend on the precise geometry. We resort to a simplified model of a purely one-dimensional infinite nanowire along $\hat x$, passing through $z=R$, where $R$ plays the role of the nanowire radius. The wire is embedded in a dielectric medium with constant $\epsilon$ that spans the whole $z>0$ half-space. This medium represents the semiconductor material of the nanowire ($\epsilon=17.7$ for InSb). For $z<0$ we assume a different dielectric with constant $\epsilon'$, representing the dielectric substrate ($\epsilon'=3.9$ for SiO${}_2$). We neglect all other elements of the experimental setup.

Classically, the bound charge density $\rho_b(\bm r)$ arising from a given density $\rho(x)$ in the nanowire can be computed exactly in this geometry using the method of images. It is a textbook result \cite{Batygin:75} that a point charge $q$ at $\bm r_q=(0,0,R)$ in a $z>0$ dielectric $\epsilon$ at a distance $R$ from another semi-infinite $z<0$ dielectric $\epsilon'$, generates an image at $\bm r_{q'}=(0,0,-R)$ with charge $q'=q(\epsilon-\epsilon')/(\epsilon+\epsilon')$. The potential $\phi_b(\bm r)$ created in the $z>0$ half-space by the physical bound charge density $\rho_b(x,y,0)$, confined at the $z=0$ interface, is equivalent to the potential created by the fictitious image $q'$ in an infinite medium with uniform dielectric constant $\epsilon$.
\[
\phi_b(\bm r)=\frac{\epsilon-\epsilon'}{\epsilon+\epsilon'}\frac{q}{4\pi\epsilon_0\epsilon}\frac{1}{|\bm r-\bm r_{q'}|}
\]
Note that if $\epsilon>\epsilon'$, this potential exerts a repulsive force on any other charge on the wire, but also on $q$ itself. Such self-interaction is physical, and is just the repulsion between $q$ and the dielectric boundary itself. As $\phi_b$ is linear in $q$ we may generalise the above result to an arbitrary linear charge density $\rho(x)$ at $\bm r=(x,0,R)$, 
\[
\phi_b(x)=\frac{\epsilon-\epsilon'}{\epsilon+\epsilon'}\frac{1}{4\pi\epsilon_0\epsilon}\int dx'\,\frac{\rho(x')}{\sqrt{(2R)^2+(x-x')^2}}
\]


In a quantum setting, where $\hat \rho$ and $\hat \rho_b$ are operators, the same problem may be tackled by assuming that bound charges and charges in the nanowire are distinguishable, and cannot tunnel between each other. This allows us to treat their interaction at a Hartree level, without any Fock correction. For an instantaneous interaction vertex between them, and assuming purely local polarizability, we may write down an equation for $\langle \hat\rho_b\rangle$ that is perfectly equivalent to the above classical equation, only changing $\rho(x')$ by $\langle\hat\rho(x')\rangle$. Within these assumptions, therefore, we can integrate out the bound charge degrees of freedom self-consistently. 
We then obtain an effective Hamiltonian for the nanowire charges of the form, 
\begin{eqnarray}\label{Hint}
H_{int}&=&\int dx\, \phi_b(x)\hat\rho(x) = \int dx\,dx'\,\langle\hat\rho(x')\rangle V_b(x'-x)\hat\rho(x)\nonumber\\
V_b(x)&=&\frac{\epsilon-\epsilon'}{\epsilon+\epsilon'}\frac{1}{4\pi\epsilon_0\epsilon}\frac{1}{\sqrt{(2R)^2+x^2}}
\end{eqnarray}
Note that the long-distance $V_b\approx 1/|x|$ Coulomb-like decay is due to the bound nature of $\rho_b$, which cannot screen out the $\hat \rho(x)$ charges. At short distances, the interaction is regularised to a finite value $V_b\propto 1/R$, which is a natural consequence of bound charges to always remain at a distance greater or equal than $R$ from nanowire charges. This behaviour is independent of geometrical details.

We emphasise that the self-interaction implicit in the above Hartree-like Hamiltonian $H_{int}$ is not an artefact of any approximation, as it arises in the classical problem as the effective interaction between nanowire charges and physical bound charges. This is unlike the case of intrinsic interactions (charge in a homogeneous infinite dielectric) discussed in the preceding section, where classically no interaction exists of an electron with itself.

\section{Incompressibility in the $V_Z-\mu$ parameter space}

In this section we present a simplified analytical description of the pattern of incompressible regions in the $V_Z-\mu$ parameter space, and derive closed expressions for relevant quantities in the thermodynamic limit of large nanowire length $L$, such as the self-consistent potential $\phi(L/2)$, the charge density $\rho(L/2)$, or the slope of incompressible regions $s$.

The central object that controls all the relevant quantities in the interacting system is the self-consistent interaction-induced potential $\phi(x)$. As shown in Fig. 2a of the main text, this potential decreases around the boundaries $x=0,L$ of the nanowire. However, what controls the overlap between MZMs and the incompressible domains is actually the \emph{bulk} value $\phi(L/2)$ at the center of the nanowire. This value is taken, by definition, as zero at $V_Z=0$ (so that the $k=0$ Fermi energy at the center of the nanowire is $\mu$ with and without interactions), and otherwise defines an effective Fermi energy at $V_Z\neq 0$, 
\begin{equation}\label{EFeff}
\varepsilon_F^\mathrm{eff}=\mu+V_Z-\phi(L/2).
\end{equation}
Note that for non-zero $V_Z$ the spin-resolved bands of the single-mode nanowire split, so that the band bottom of each lies at $\mu\pm V_Z-\phi(L/2)$ (for small $\Delta$). Equation \eqref{EFeff} then corresponds to the Fermi energy of the deeper subband which, unlike its companion, remains populated in the topological regime.

While the Fermi energy $\varepsilon_F^\mathrm{eff}$ above corresponds to the bulk $L\to \infty$ nanowire, it also allows to accurately describe finite-$L$ nanowires as long as the typical level spacing 
\[
\delta\epsilon=\frac{\hbar^2}{2mL^2}
\]
is smaller than $\varepsilon_F^\mathrm{eff}$ ($\delta\epsilon=2.5\mu$eV for the $1\mu$m InSb nanowire in the main text). We therefore derive an approximate solution for $\phi\equiv\phi(L/2)$ (uniform potential $\phi(x)$) in the thermodynamic limit $L\to\infty$. This is possible by taking the pairing $\Delta$ and spin-orbit $\alpha$ as weak perturbations. In that case, the (uniform) charge density of the infinite nanowire at a given $\mu,V_Z$ is the sum of the two parabolic, Zeeman-split subbands,
\begin{equation}\label{rho}
\rho=\frac{2e}{\pi L}\mathrm{Re}\left(\sqrt{\frac{\mu+V_Z-\phi}{\delta\epsilon}}+\sqrt{\frac{\mu-V_Z-\phi}{\delta\epsilon}}\right)
\end{equation}


The self-consistency equation for repulsive interactions $V_b$ with bound charges reads
\begin{equation}\label{SCE}
\phi=L	\bar V_b\left(\rho-\rho_0\right),
\end{equation}
where
\begin{eqnarray*}
\bar V_b&=&\frac{1}{L}\int_{-L/2}^{L/2} V_b(x,0)>0 \\
\rho_0&=& \left.\rho\right|_{V_Z=0}= 2\frac{2e}{\pi L} \mathrm{Re}\sqrt{\frac{\mu}{\delta\epsilon}}
\end{eqnarray*}

We are interested in the solution to Eq. \eqref{SCE} in the topological regime, $V_Z>V_Z^c\approx\mu$ (recall that $\Delta$ is a small perturbation), so that the second term in Eq. \eqref{rho} is zero. This allows for a simple analytical solution in this regime,
\begin{equation}\label{phisol}
\phi=\frac{1}{2}\left(\sqrt{v^2+4v\left(V_z+\mu+2\mathrm{Re}\sqrt{v\mu}\right)}-4\mathrm{Re}\sqrt{v\mu}-v\right),
\end{equation}
where $v=4 \bar V_b^2/(\pi^2\delta\epsilon)$ is an energy scale measuring the strength of the interactions relative to the level spacing of the nanowire ($v\approx 9.2$ meV for the nanowire in the main text).

Note that the solution Eq. \eqref{phisol} is quite different for $\mu>0$ and $\mu<0$, due to the $\sqrt{\mu}$ branch cuts. For example, if we evaluate the difference in charge density at the transition and at zero magnetic field, $\rho_{V_Z=\mu}-\rho_0=\phi_{V_Z=\mu}/(L\bar V_b)$, we find it is zero for $\mu<0$, but negative for $\mu>0$, as obtained numerically in Fig. 2c of the main text.

The incompressible regions in the interacting phase diagram, Fig. 4b of the main text, correspond to contours where $\varepsilon_F^\mathrm{eff}$ in Eq. \eqref{EFeff} is constant. This follows from the fact that parity crossings in a non-interacting finite-$L$ nanowire occur at integer values of $L/\lambda_F$, where $\lambda_F$ is the $\Delta=0$ Fermi wavelength \cite{Klinovaja:PRB12}. Therefore, since pinned regions emerge around parity crossings, they follow the constant $\varepsilon_F^\mathrm{eff}$ contours. This reasoning allows us to derive an expression for the slope $s$ of the incompressible regions throughout the $V_Z-\mu$ plane,
\[
s=-\frac{\partial\varepsilon_F^\mathrm{eff}/\partial V_Z}{\partial\varepsilon_F^\mathrm{eff}/\partial \mu}=-\frac{1-\partial\phi/\partial V_Z}{1-\partial\phi/\partial \mu}
\]
Note that this is the ratio between the $L\to\infty$ $\mu$- and $V_Z$-compressibilities $\kappa_\mathrm{eff}$ and $g_\mathrm{eff}/g$. Note also that since $Q_M$ charge jumps are averaged out in the $L\to\infty$ limit (Fig. 2d, main text), these compressibilities are actually averaged values of the rapidly varying ones for finite $L$ (Fig. 3b, main text).

Inserting the solution of Eq. \eqref{phisol} into the expression for the slope $s$, we find  
\begin{eqnarray*}
s=\left\{\begin{array}{cc}
-1 &\textrm{for $\mu<0$} \\
-\left(1+\sqrt{v/\mu}\right)^{-1}&\textrm{for $\mu>0$} 
\end{array}\right.,
\end{eqnarray*}
which closely follows the numerical results in Fig. 4b of the main text. The reduced slope $-1<s<0$ for $\mu>0$ is physically due to a suppression of $\kappa_\mathrm{eff}$ relative to $g_\mathrm{eff}/g$, and can be connected to the negative $\rho_{V_Z=\mu}-\rho_0$ induced by the interactions at finite nanowire densities $\rho_0$.

\bibliography{biblio}

\end{document}